\documentclass[aps,prd,twocolumn,superscriptaddress,showpacs,amsmath,amssymb]{revtex4}
\usepackage{comment}
\usepackage{multirow}
\usepackage{dcolumn}% Align table columns on decimal point
\usepackage{bm}% bold math
\usepackage{setspace}
\usepackage{graphicx}% Include figure files
\usepackage{xspace}

\usepackage{dcolumn}% Align table columns on decimal point
\usepackage{bm}% bold math
\newenvironment{cfigure1c}[1][tbp]{\begin{figure*}[#1]\centering}{\end{figure*}}

\newcommand{\ppbar}  {\ensuremath{p\bar{p}}\xspace}
\newcommand{\ttbar}  {\ensuremath{t\bar{t}}\xspace}

\newcommand{\et}     {\ensuremath{E_{T}}\xspace}

\newcommand{\pt}     {\ensuremath{p_{T}}\xspace}

\newcommand{\met}    {\ensuremath{\text{\raisebox{.3ex}{$\not$}}\et}\xspace}

\newcommand{\chisq}  {\ensuremath{\chi^{2}}\xspace}

\newcommand{\genunit}[2]{\ensuremath{#1~\mathrm{#2}}\xspace}

\newcommand{\gev}[1]    {\genunit{#1}{GeV}}

\newcommand{\gevc}[1]   {\ensuremath{#1~\mathrm{GeV}/c}}
\newcommand{\gevcc}[1]  {\ensuremath{#1~\mathrm{GeV}/c^{2}}}

\newcommand{\invfb}[1]  {\ensuremath{#1~\mathrm{fb}^{-1}}}

\begin{document}

%\setpagewiselinenumbers
%\modulolinenumbers[1]
%\linenumbers

\title{Azimuthal decorrelation in  \ttbar production at hadron colliders}
\affiliation{Department of Physics, Korea University, Seoul 136-713, Korea}
\affiliation{Department of Physics, Ewha Womans University, Seoul, 120-750, Korea}
\author{Suyong Choi}
\email[]{suyong@korea.ac.kr}
\affiliation{Department of Physics, Korea University, Seoul 136-713, Korea}
%\email{suyong@korea.ac.kr}
\author{Hyun Su Lee}
\email[Corresponding Author: ]{hyunsulee@ewha.ac.kr}
\affiliation{Department of Physics, Ewha Womans University, Seoul, 120-750, Korea}
%\email{hyunsulee@ewha.ac.kr}
%\affiliation[a]{Department of Physics, Korea University, Seoul 136-713, Korea}
%\affiliation[b]{Department of Physics, Ewha Womans University, Seoul, 120-750, Korea}

%\emailAdd{suyong@korea.ac.kr}
%\emailAdd{hyunsulee@ewha.ac.kr}

\date{\today}% It is always \today, today,
\begin{abstract}
We present a new observable, $\Delta \phi$, an azimuthal angle difference between $t$ and $\bar{t}$ quarks in  \ttbar pair production, at  hadron colliders as an interesting probe of the radiative quantum chromodynamics process as well as a high-order correction in the high-mass regime. This variable also enables good discrimination on some new physics models that may explain the forward-backward charge asymmetry of \ttbar production measured at the Tevatron. With a reliable estimation of the dataset obtained up to 2011 at the Tevatron and Large Hadron Collider, we present an opportunity for testing the standard model as well as searching new physics models with the $\Delta \phi$ observable in the lepton+jets decay channel.
\end{abstract}

% PACS, high energy physics
%\keywords{QCD, \ttbar, Beyond Standard Model}%Use showkeys class option if keyword
%\begin{document}
\maketitle
%\flushbottom

\section{Introduction}
Azimuthal correlation of high transverse momentum~(\pt) jets is a valuable probe of the predictive power of quantum chromodynamics~(QCD) and allows one to study QCD radiative processes. An accurate description of QCD radiative processes is crucial for a wide range of hadron collider measurements, including precision tests of the standard model~(SM) as well as discoveries of new particles such as the Higgs boson and supersymmetry~(SUSY) partners.
In hadron colliders, the most common type of event is two-jet production with  equal energies transverse to the beam direction such that the two jets are correlated in the azimuthal angle~($\phi$) and the difference between the azimuthal angles of the two jets~($\Delta \phi$) is equal to $\pi$. However, additional particles or jets including low-energy particles produced in the same event reduce $\Delta \phi$ to less than $\pi$. Therefore, the measurement of azimuthal decorrelation provides an ideal test to understand the hard and soft radiative processes of QCD. Because of their importance,  azimuthal decorrelations have been widely studied with inclusive two-jet production in  various hadron collider experiments~\cite{d0dijet,cmsdijet,atlasdijet}.

Understanding heavier quark  production is particularly important because most new physics particles, which are usually quite heavy, decay into heavy quarks in the  SM. The CDF Collaboration has  measured the azimuthal angle decorrelation of bottom~($b$) quarks from  $b\bar{b}$ pair production~\cite{cdf_bbbar}. However, so far, there has been no measurement of the azimuthal decorrelation of top~($t$) quarks, which are the heaviest known elementary particles. Measurement of  $\ttbar$ pair production and the radiative jet process is quite important for understanding higher order QCD effects with the heaviest elementary particle. Understanding of $\ttbar$ pair production have been done in hadron colliders with various differential cross section measurements for interesting observables such as invariant mass of \ttbar, $\pt$ of \ttbar, and $\pt$ of top quarks so on~\cite{diff_cross}, but not for the azimuthal decorrelation. The measurement of the azimuthal angle decorrelation, therefore, can be a complementary test of the SM in the \ttbar system.

Measurements of the top quark had been limited by the relatively small cross section of \ttbar at the Tevatron, with an order of 1,000 events being listed in their full dataset. However, the Large Hadron Collider~(LHC) has already obtained approximately \invfb{5} $pp$ collisions at $\sqrt{s}=7$~TeV, corresponding to an order of 10,000 \ttbar events, by taking advantage of the approximately 20 times larger production cross section. With large statistics of  \ttbar events at the LHC, we can perform precision studies of the SM processes in the top sector. Even with limited statistics at the Tevatron, it is still interesting to study the feasibility of  $\Delta \phi$ measurement under  different initial conditions using the full Tevatron dataset.

The top quark, the most recently discovered quark (in 1995~\cite{ttbar_observe} at the Fermilab Tevatron \ppbar Collider), is the heaviest known elementary particle. Its large mass may indicate strong coupling with electroweak symmetry breaking, and therefore, the top quark is usually treated differently from the other light quarks in many new physics models. This suggests that many searches focus on the top quark signature. The recent observation of the charge forward-backward asymmetry at the Tevatron~\cite{cdf_afb,d0_afb} may be evidence for a new physics signature involved in  \ttbar production.
However, it seems hard to confirm the observation of new physics with a significance of 5 standard deviation from the SM with the limited statistics of Tevatron data even with the full dataset of  10~fb$^{-1}$~\cite{tev_afbfull}. Since the LHC is a $pp$ collider, it is difficult to probe all the possible scenarios of the Tevatron charge asymmetry at the LHC. It is even more difficult to distinguish the most relevant theories and parameters with the charge asymmetry measurement alone.
The azimuthal decorrelation between $t$ and $\bar{t}$ quarks in  \ttbar pair production is sensitive to additional radiation as well as the production mechanisms. Therefore, we expect it to be sensitive to new physics models and believe that it may provide an additional discrimination among the models related to forward-backward charge asymmetry at the Tevatron.
In this paper, we propose a new observable, $\Delta \phi$, the azimuthal angle between $t$ and $\bar{t}$ quarks in \ttbar pair production at the hadron colliders,  as a general probe of the dynamics of \ttbar production.

\section{Truth level comparison of different models }

\begin{cfigure1c}
\begin{tabular}{cc}
\includegraphics[width=0.43\textwidth]{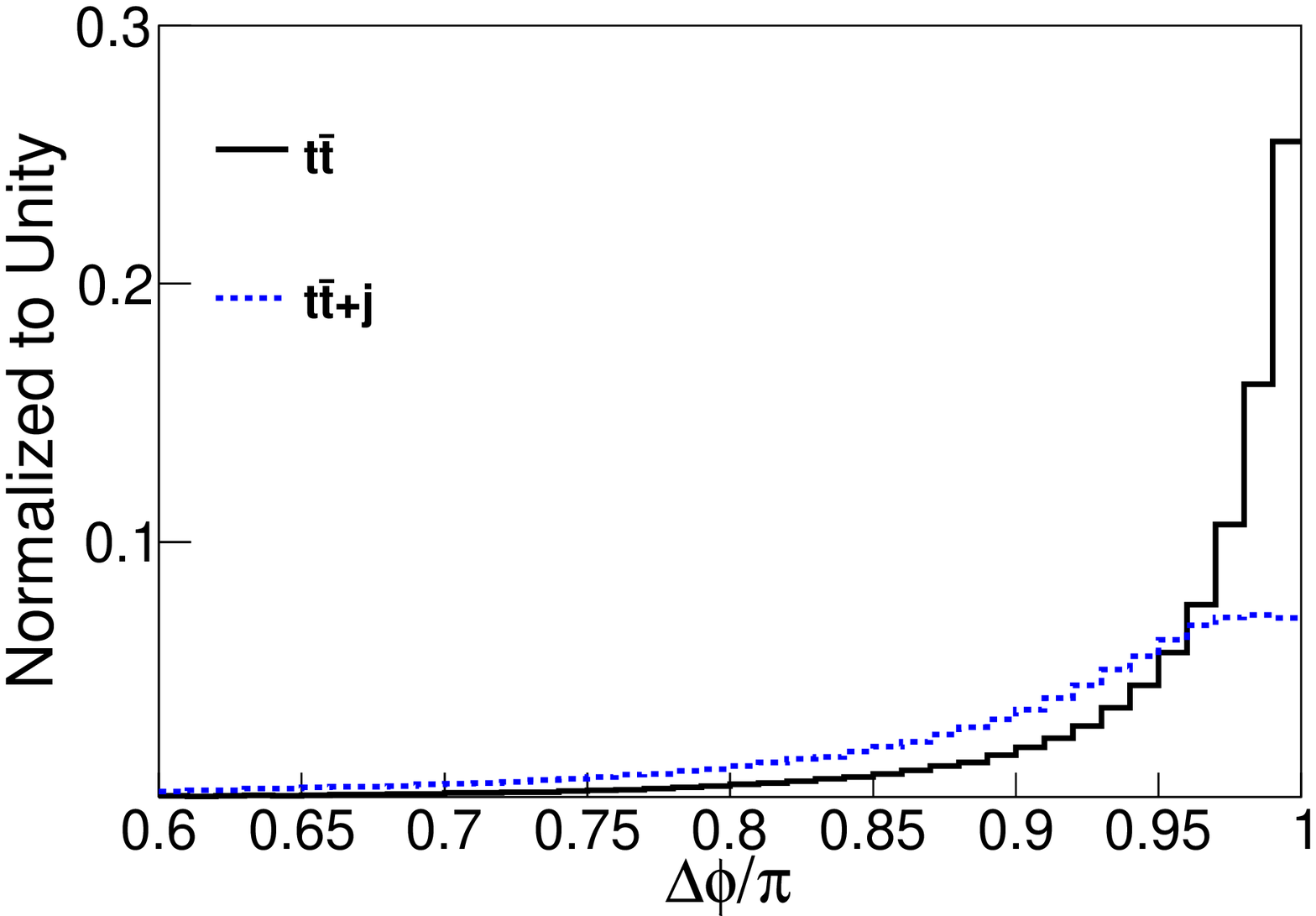} &
\includegraphics[width=0.43\textwidth]{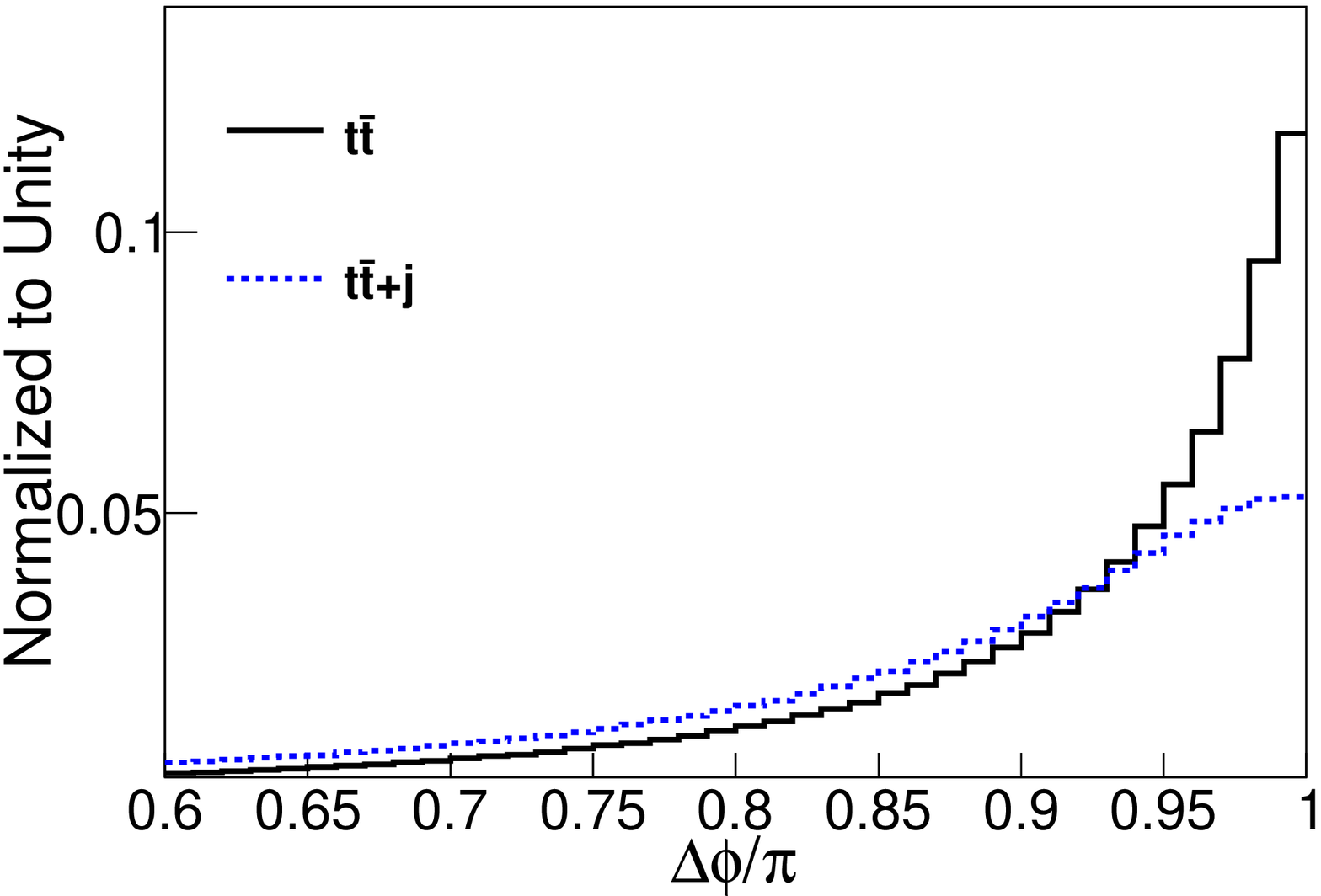} \\
(a) Tevatron  & (b) LHC \\
\end{tabular}
\caption[Data fit]{(Color online) Truth level comparison of azimuthal angle between $t$ and $\bar{t}$ between $\ttbar$ and $\ttbar j$ samples at the Tevatron and LHC.}
\label{ref:hep}
\end{cfigure1c}

\begin{cfigure1c}
\begin{tabular}{cc}
\includegraphics[width=0.43\textwidth]{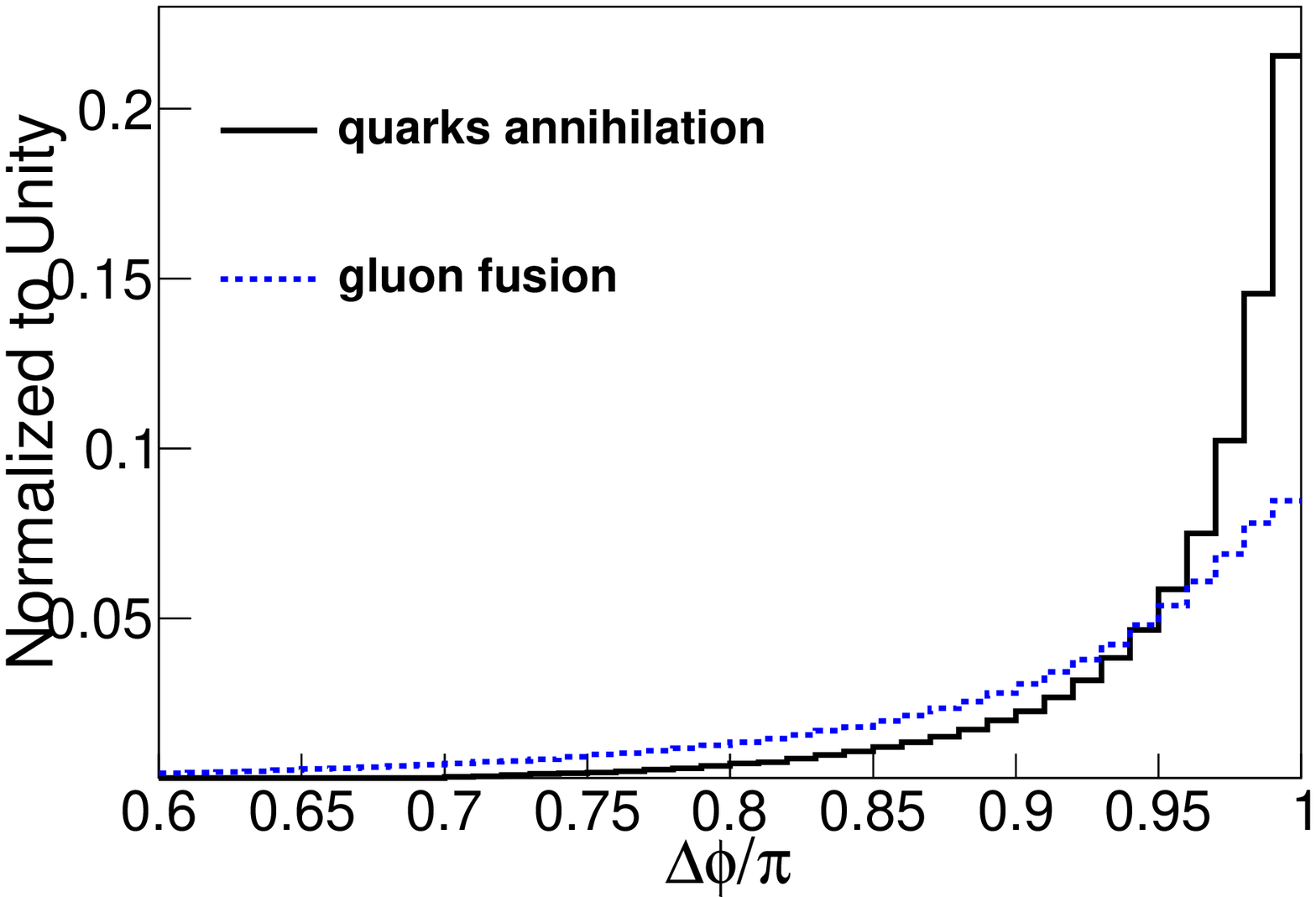} &
\includegraphics[width=0.43\textwidth]{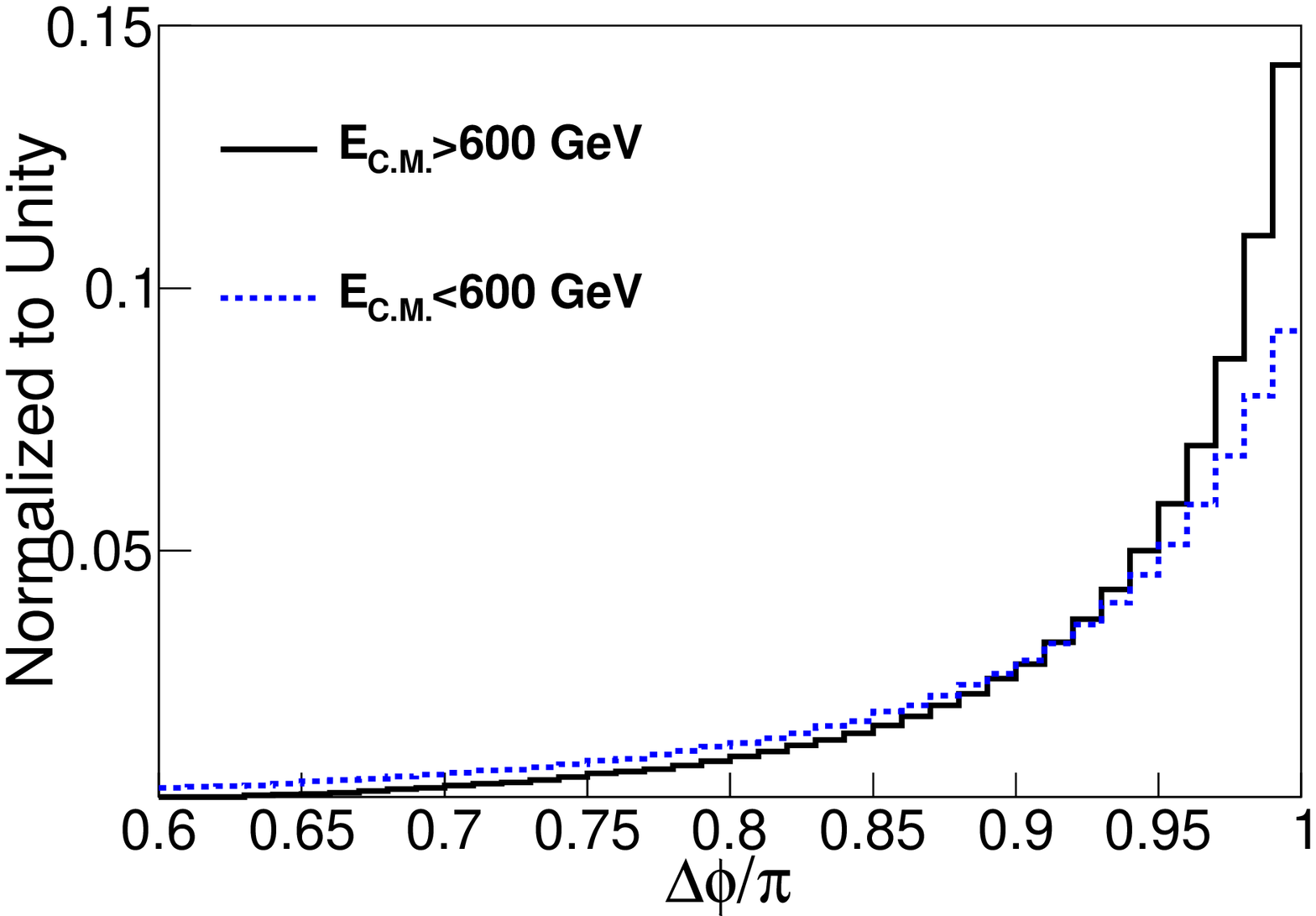} \\
(a) Production mechanism  & (b) $E_{\text{C.M.}}$ \\
\end{tabular}
\caption[Data fit]{(Color online) Truth level comparisons of the azimuthal angle between $t$ and $\bar{t}$ with different states of the collision partons.}
\label{ref:inparton}
\end{cfigure1c}

\begin{cfigure1c}
\begin{tabular}{cc}
\includegraphics[width=0.43\textwidth]{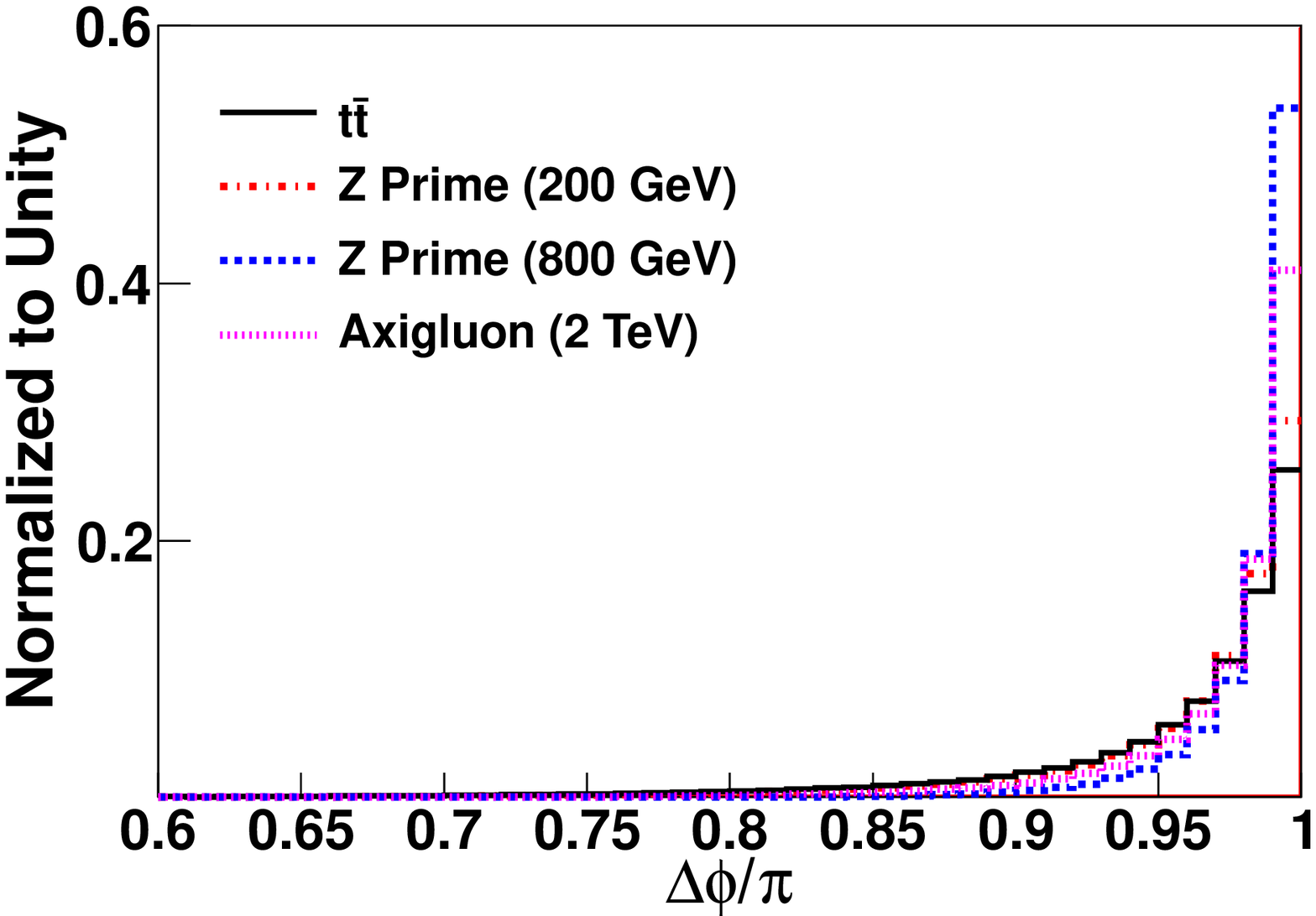} &
\includegraphics[width=0.43\textwidth]{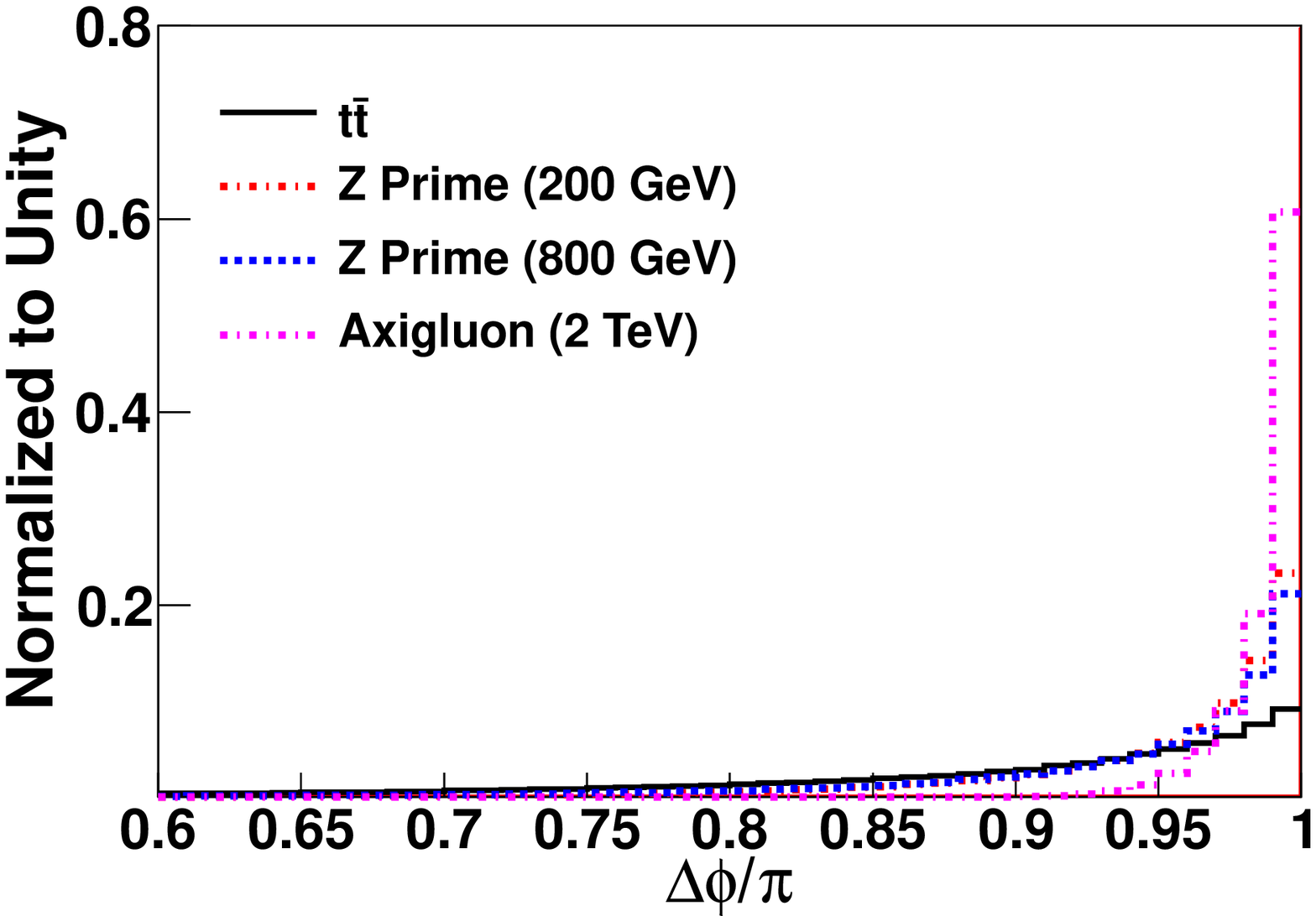} \\
(a) Tevatron (New Physics) & (b) LHC (New Physics)\\
\includegraphics[width=0.43\textwidth]{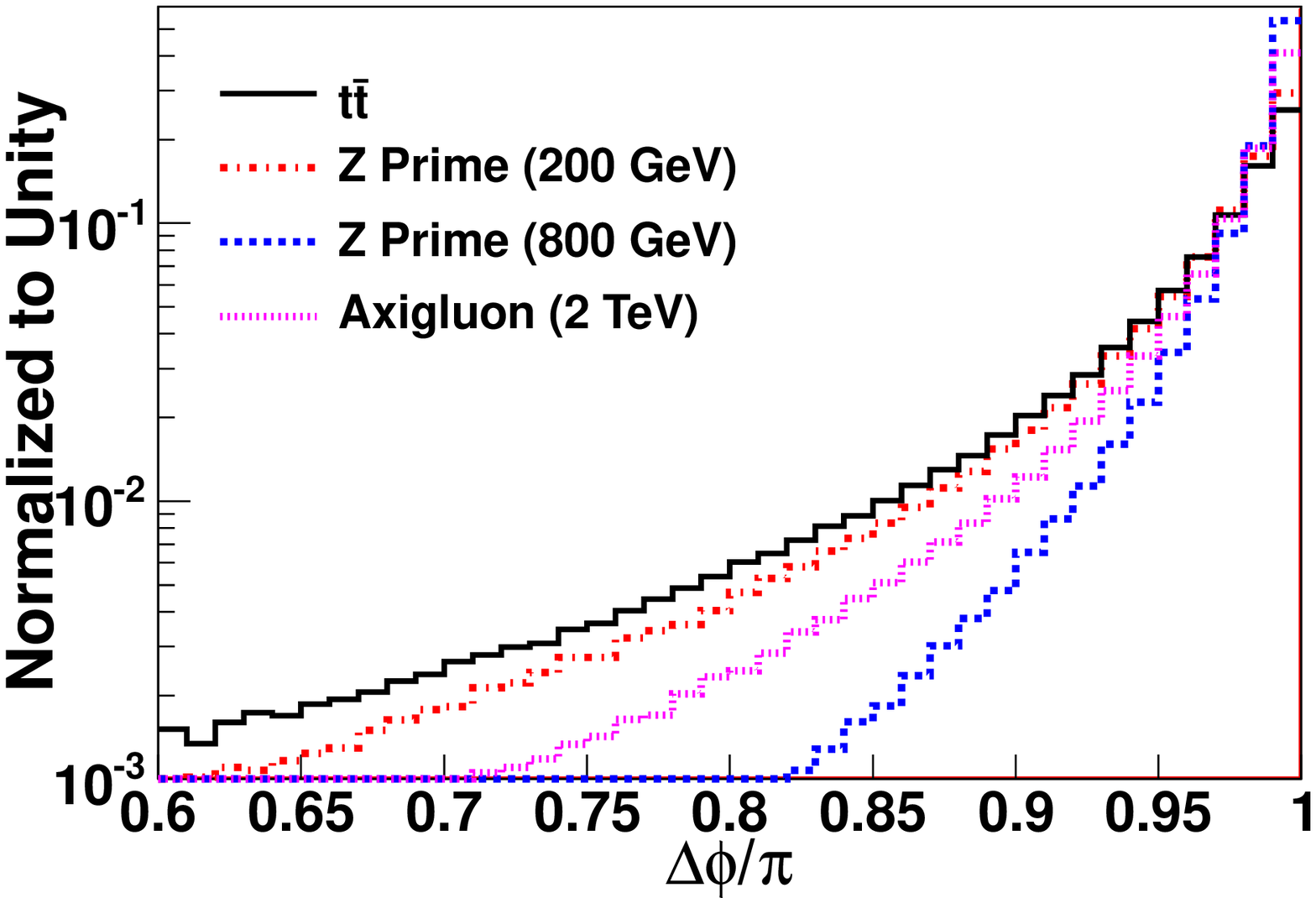} &
\includegraphics[width=0.43\textwidth]{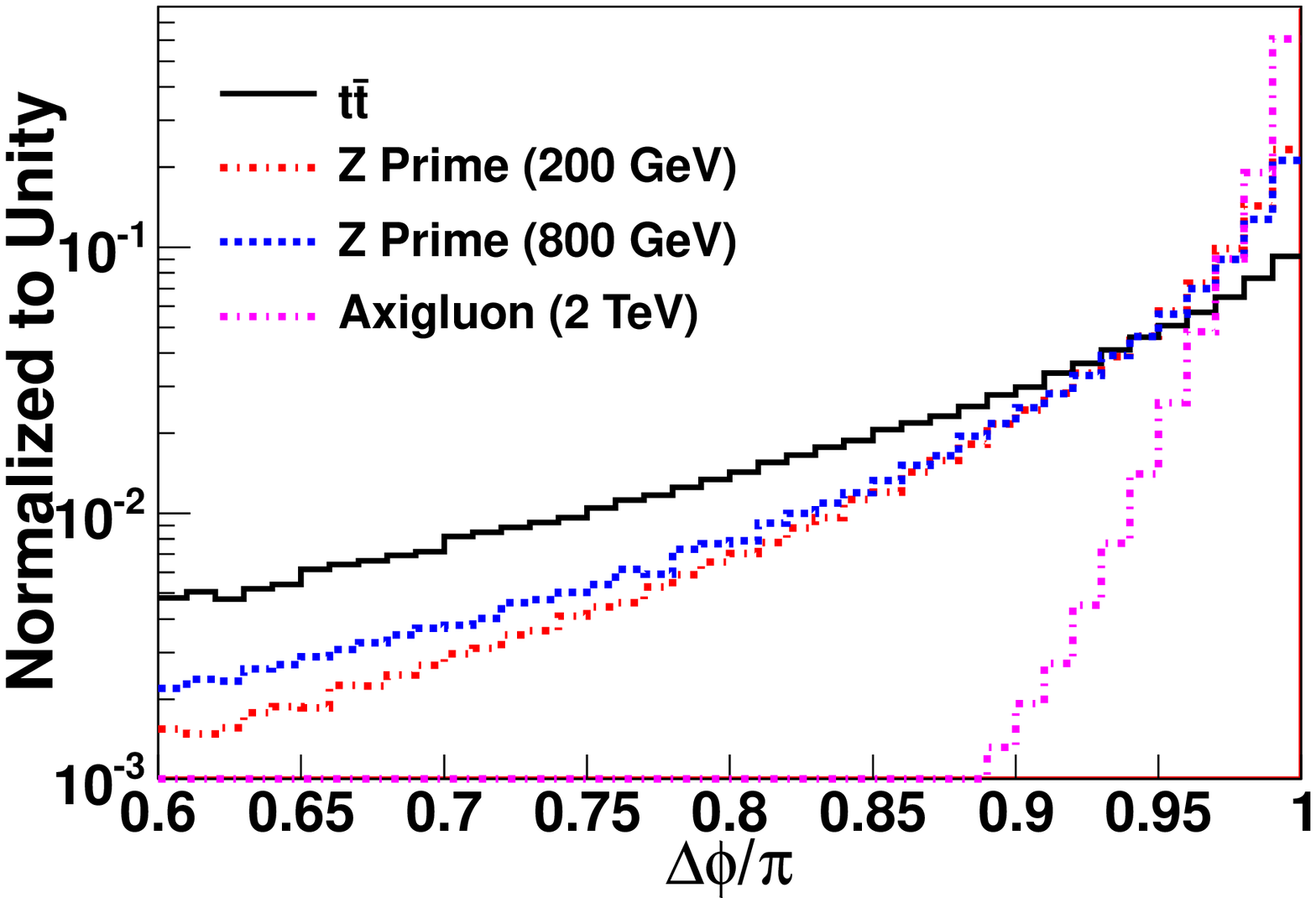} \\
(c) Tevatron (Log Scale)& (d) LHC (Log Scale)\\
\end{tabular}
\caption[Data fit]{(Color online) Truth level comparisons of the azimuthal angle between $t$ and $\bar{t}$ of various new physics samples at the Tevatron and LHC.}
\label{ref:hep_np}
\end{cfigure1c}

We generated simulated \ttbar signal samples using the leading order~(LO) Monte Carlo~(MC) generator {\sc madgraph/madevent} package~\cite{madgraph} with {\sc pythia} parton showering~\cite{pythia}. 
We generated signal samples for the inclusive production of \ttbar as well as a \ttbar with an additional hard jet~($\ttbar j$). We request the extra jet having transverse momentum~(\pt) greater than \gevc{20}. Here most of additional jets in the $\ttbar j$ samples were originated from initial- or final-state gluon radiations so that $\ttbar j$ is the radiation-enriched sample. 
We generated 400,000 events in each sample for the Tevatron and 2,000,000 events for the LHC. To verify the distribution of $\Delta \phi$ values from different processes, we generated plots comparing \ttbar with $\ttbar j$ samples, as shown in Fig.~\ref{ref:hep}. In these plots, we use truth level information of four vectors from $t$ and $\bar{t}$ including initial state soft radiation. As already discussed, $\Delta \phi$ is a variable sensitive  to the radiation under conditions where the shapes corresponding to \ttbar and $\ttbar j$ are very different. 

There are a couple of interesting models beyond the standard model~(BSM) that can participate in  \ttbar production. There are several models that can explain the charge asymmetry at the Tevatron. For color-singlet heavy bosons, such as Z$'$, mediated production is one of the most interesting scenarios~\cite{heavyZP}. However, the measured constraints on Z$'$ from dijet production at the LHC already surpass the TeV scale~\cite{LHC_Zprime}. Also, a heavy resonant Z$'$ cannot explain the total cross section of \ttbar at the Tevatron, which is in excellent agreement with that of the SM. To avoid this issue,  low-mass-vector-boson-mediated production is considered~\cite{lightZP}. A direct search for low-mass vector bosons, such as Z$'$, is very difficult because of the low production cross section as well as large SM backgrounds. For our study of the $\Delta \phi$ observable, we generated simulated signal samples for low-mass,  nonresonant \ttbar production from 200-GeV Z$'$ bosons. We also generated samples for the decay of 800-GeV resonant Z$'$ production into \ttbar as a benchmark model. We consider massive color-octet vector boson production, for which there are well-motivated theories to explain the charge asymmetry at the Tevatron~\cite{afb_theory}, such as axigluon models~\cite{axigluon}, technicolor models~\cite{technicolor}, and extra-dimensional models with KK gluons~\cite{kkgluon}. Even though the LHC has started investigating dijet resonance in the TeV scale, it is not yet able to probe the color-octet vector bosons associated with the electroweak symmetry breaking sector, as suggested by several of the models discussed above. As a benchmark model for color-octet vector boson production, we generated signal samples for 2-TeV axigluon-mediated production decaying into \ttbar.
We used the {\sc madgraph/madevent} package with the top-BSM model~\cite{topBSM} to generate these new physics models.

In the truth level, the azimuthal decorrelation is only affected by the transverse components of the initial state radiations~(ISRs). Therefore the states of collision partons are quite important. The \ttbar production through quark and antiquark annihilation~($q\bar{q}$) usually have less ISR than the production through gluon fusion~($gg$). Even though the SM \ttbar can be produced via $gg$ and $q\bar{q}$, the new physics processes are produced mostly via $q\bar{q}$. Another important thing related with the azimuthal angle decorrelation is the center of mass energy~($E_{C.M.}$) of the two colliding partons. The new physics processes mediated with heavy particle definitely have much larger $E_{C.M.}$ than the process from the SM. Radiation from higher energy parton is relatively collinear with initial parton. Since the colliding partons only have longitudinal momenta, ISRs from higher energy collision partons  have smaller transverse momentum. To verify this effect, we study the collision partons of the SM \ttbar production. In Fig.~\ref{ref:inparton} (a), we show the $\Delta \phi$ distributions from $gg$ and $q\bar{q}$ productions. We also make a plot in Fig.~\ref{ref:inparton} (b) with different criteria of $E_{C.M.}$ requirements, $E_{C.M.}>\gev{600}$ and $E_{C.M.}<\gev{600}$. These results clearly show the effects from the collision partons.

Because the initial state radiations depend on the initial partons as well as the mediated particles of \ttbar production, $\Delta \phi$ distributions from different new physics models can be different with the SM distribution.
We compare the $\Delta \phi$ distribution from new physics models with the SM \ttbar in Fig.~\ref{ref:hep_np} using truth information with the initial state soft radiation. The new physics models are clearly separable from the SM \ttbar in  most of the models for both the Tevatron and the LHC. At the Tevatron,  heavy Z$'$ production exhibits the most visible difference with the SM \ttbar, and the 2-TeV axigluon production is also   clearly distinguished. However, $\Delta \phi$ distribution with the 200-GeV Z$'$ production shows relatively small deviation from the SM \ttbar, which indicates that we may need larger event statistics to separate this new physics model from the SM.  At the LHC, the difference is more marked. The three new physics models are clearly separated from the SM. In addition, we can take advantage of the large cross section at the LHC, which would  provide us a much better opportunity to detect new physics with $\Delta \phi$.  We investigated, in detail, the realistic situation after carrying out detector simulations as well as event reconstruction. We assume the number of events to be \invfb{10} at $\sqrt{s} = 1.96 $TeV Tevatron and \invfb{5} at $\sqrt{s} = 7 $ TeV LHC.

\section{Models with detector simulation}
It is important to consider the effect of detector resolution and data statistics  as well as the background contributions for a realistic prediction of the measurement. The parton level information of each model generated with {\sc madgraph/madevent} and {\sc powheg} has undergone parton showering and hadronization using {\sc pythia}. The detector effects are taken into account by using the fast detector simulation package~{\sc pgs}~\cite{pgs}. The detector resolution effects are simulated using the following parameterization:
$$\frac{\delta E}{E} = \frac{a}{\sqrt{E}}~ ~ \mathrm{for~ jets,}$$
$$\frac{\delta E}{E} = \frac{b}{\sqrt{E}} \oplus c~ ~ \mathrm{for~ leptons.}$$
As per the predefined values in the {\sc pgs} package, we considered $a=0.8$, $b=0.2$, and $c=0.01$ for the Tevatron and $a=1.25$, $b=0.03$, and $c=0.01$ for the LHC. The
{\sc pgs} package can also quickly reconstruct each physics object such as leptons, jets, and missing transverse energy. Jets originating from $b$ quarks are tagged with approximately 40\% $b$-tagging efficiency. 

In the SM, the top quark decays almost exclusively into a $W$ boson and a $b$ quark~\cite{pdg}. In the \ttbar events, the case where one $W$ decays leptonically into an electron or a muon plus neutrino and the other $W$ decays hadronically into a pair of jets defines the lepton+jets decay channel. Events in this channel thus contain one charged lepton, two $b$ quark jets, two light flavor quark jets, and one undetected neutrino. To select the candidate events of \ttbar lepton+jets channel, we require one charged lepton candidate with $p_{T}>\gevc{20}$. We also require a missing transverse energy exceeding \gev{20} and at least four jets with $E_{t}>\gev{20}$. We request at least one jet to be tagged as $b$ quark.

The expected signal and background events with one or more  $b$-tagged jets are taken from the Tevatron \invfb{4.8} measurement~\cite{tevnum} and the LHC \invfb{1.1} measurement~\cite{lhcnum}. The expected events with Tevatron \invfb{10} and LHC \invfb{5} are scaled based on the luminosity increase. Background events are also modeled with the {\sc madgraph/madevent} package.  To simplify the analysis, we only consider the major background, from $W$+jets, as the shape of the full background. Table~\ref{table_background} shows the expected signal and background events at the Tevatron \invfb{10} and LHC \invfb{5}.

\begin{table}
\begin{center}
\caption[Expected numbers of background and signal events]{
        Expected numbers of signal and background events at the Tevatron with 10~fb$^{-1}$ \ppbar collisions and at the LHC with 5~fb$^{-1}$ $pp$ collisions.
}
\begin{ruledtabular}
\label{table_background}
\begin{tabular}{lcc}
\ttbar signal& 1859$\pm$ 189 &46527 $\pm$ 1173 \\
Background   & 516$\pm$ 110 & 11459 $\pm$ 2345 \\
\end{tabular}
\end{ruledtabular}
\end{center}
\end{table}

The reconstruction of \ttbar events from final state particles is particularly important for estimating $\Delta \phi$. In the lepton+jets final state, the top quark momenta and neutrino momentum are fully reconstructed because  the system is overconstrained by the well-known $W$ boson mass of \gevcc{80.4}~\cite{wmass} and the $t$ quark mass of \gevcc{173}~\cite{mtop}. However, the ambiguity of jets-to-partons assignments introduces complications in event reconstruction and results in a smearing of the distribution. To obtain the most probable combination as well as to calculate the neutrino momentum, we build a $\chi^2$-like kinematic fitter. The form of the kinematic fitter used in this reconstruction is very similar to that used in the CDF measurements~\cite{cdf_fitter}. However, because of the lack of raw detector information in the fast simulation, we directly use $\met$ instead of the unclustered energy with the conservative assumption of approximately 40\% resolution. We then define $\chi^{2}$ for the kinematic fitter as
\begin{eqnarray}
\label{eq_chi2}
\chisq & = &
\Sigma_{i = \ell, 4 jets} {(p_T^{i,fit} - p_T^{i,meas})^2 / \sigma_i^2} \nonumber \\
& + & \Sigma_{k = x,y} {(\nu_{T_k}^{fit} - {\mathrm{\raisebox{.3ex}{$\not$}}E}_{T_k}^{meas})^2 / \sigma_k^2} \nonumber \\
& + & {({M}_{jj} - {M}_{W})^2 / \Gamma_{W}^2}
 + {({M}_{\ell\nu} - {M}_{W})^2 / \Gamma_{W}^2} \nonumber \\
& + &  {\{{M}_{bjj} - {M}_{\text{top}}\}^2 / \Gamma_t^2} \nonumber \\
& + & {\{{M}_{b\ell\nu} - {M}_{\text{top}}\}^2 / \Gamma_t^2}\nonumber.
\end{eqnarray}
In this $\chi^2$ formula, the first term constrains  $\pt$ of the lepton and the four leading jets to their measured values within their uncertainties~(detector resolutions);
the second term does the same for both transverse components of  \met, $x$ and $y$, as well as those of the neutrino, $p_{x}$ and $p_{y}$. The remaining four terms, the quantities
${M}_{jj}, {M}_{\ell\nu}, {M}_{bjj}$, and ${M}_{b\ell\nu}$,
refer to the invariant masses of the four-vector sum of the particles denoted in the subscripts. ${M}_W$ and ${M}_{\text{top}}$ are the masses of the $W$ boson~(\gevcc{80.4})~\cite{pdg} and
the $t$ quark~(\gevcc{173.0})~\cite{mtop}, respectively.
$\Gamma_W$ (\gevcc{2.1}) and $\Gamma_t$ (\gevcc{1.5}) are the total widths of the $W$ boson
and the $t$ quark, respectively~\cite{pdg}.

\begin{cfigure1c}
\begin{tabular}{cc}
\includegraphics[width=0.43\textwidth]{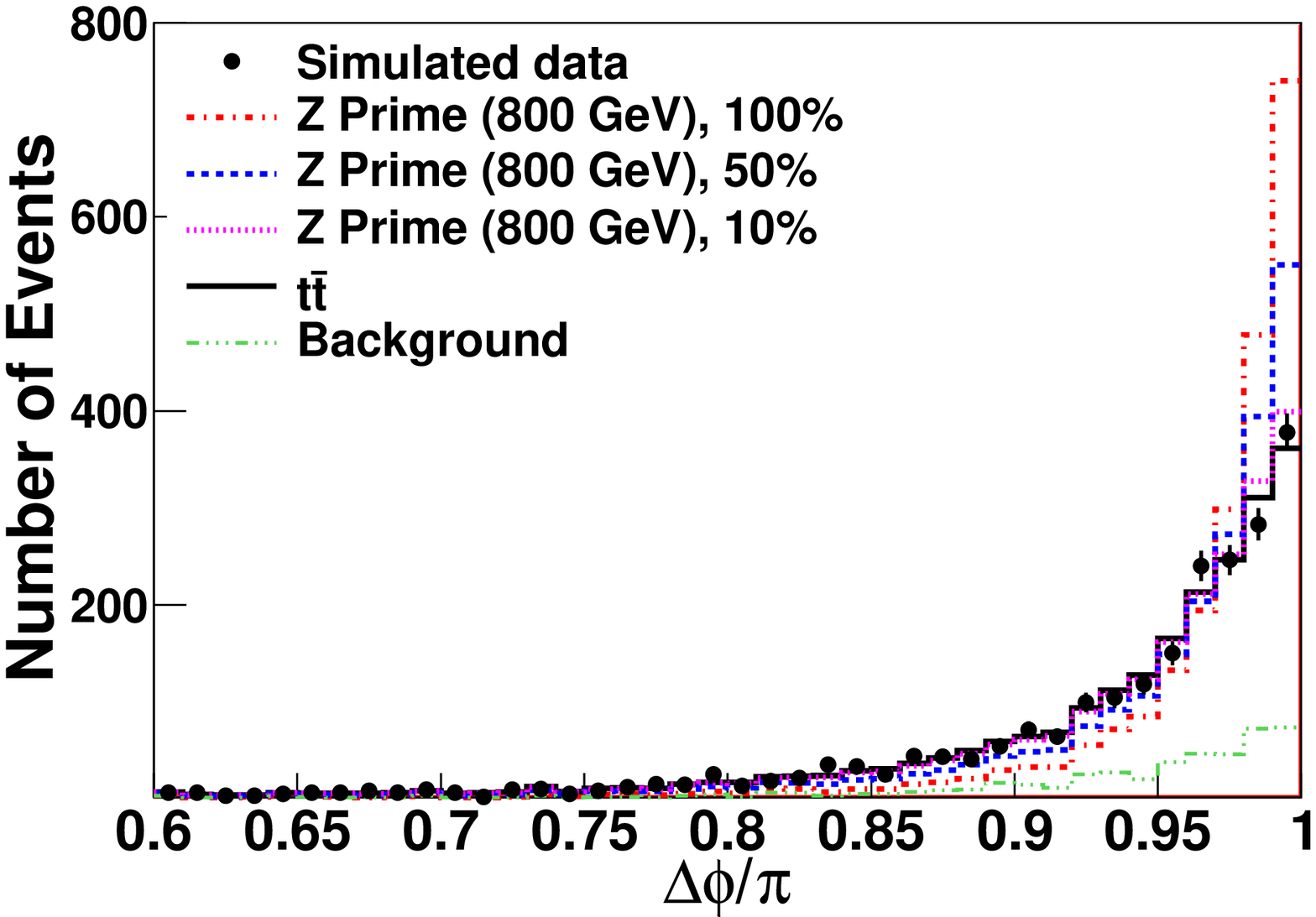} &
\includegraphics[width=0.43\textwidth]{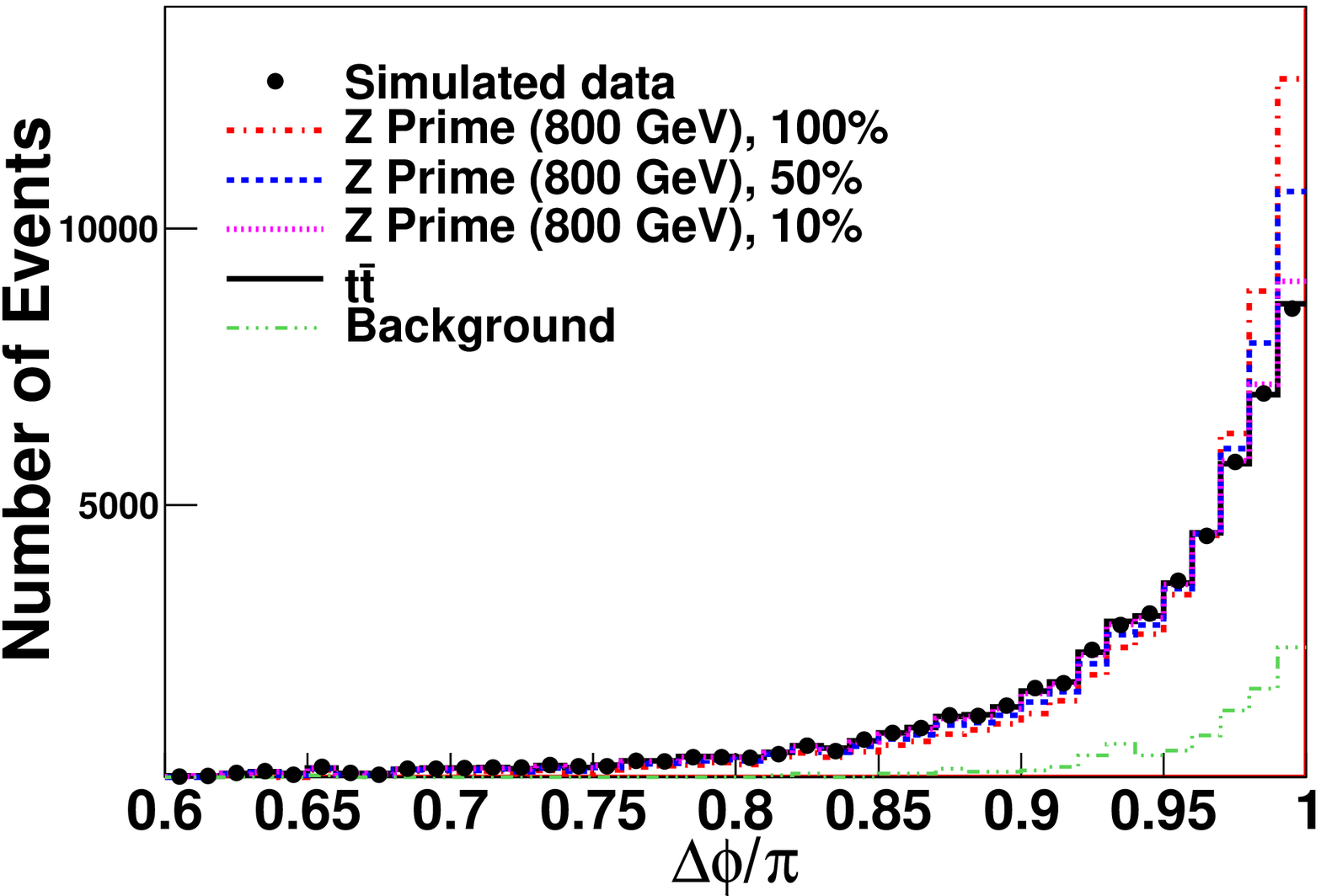} \\
(a) Tevatron & (b) LHC  \\
\includegraphics[width=0.43\textwidth]{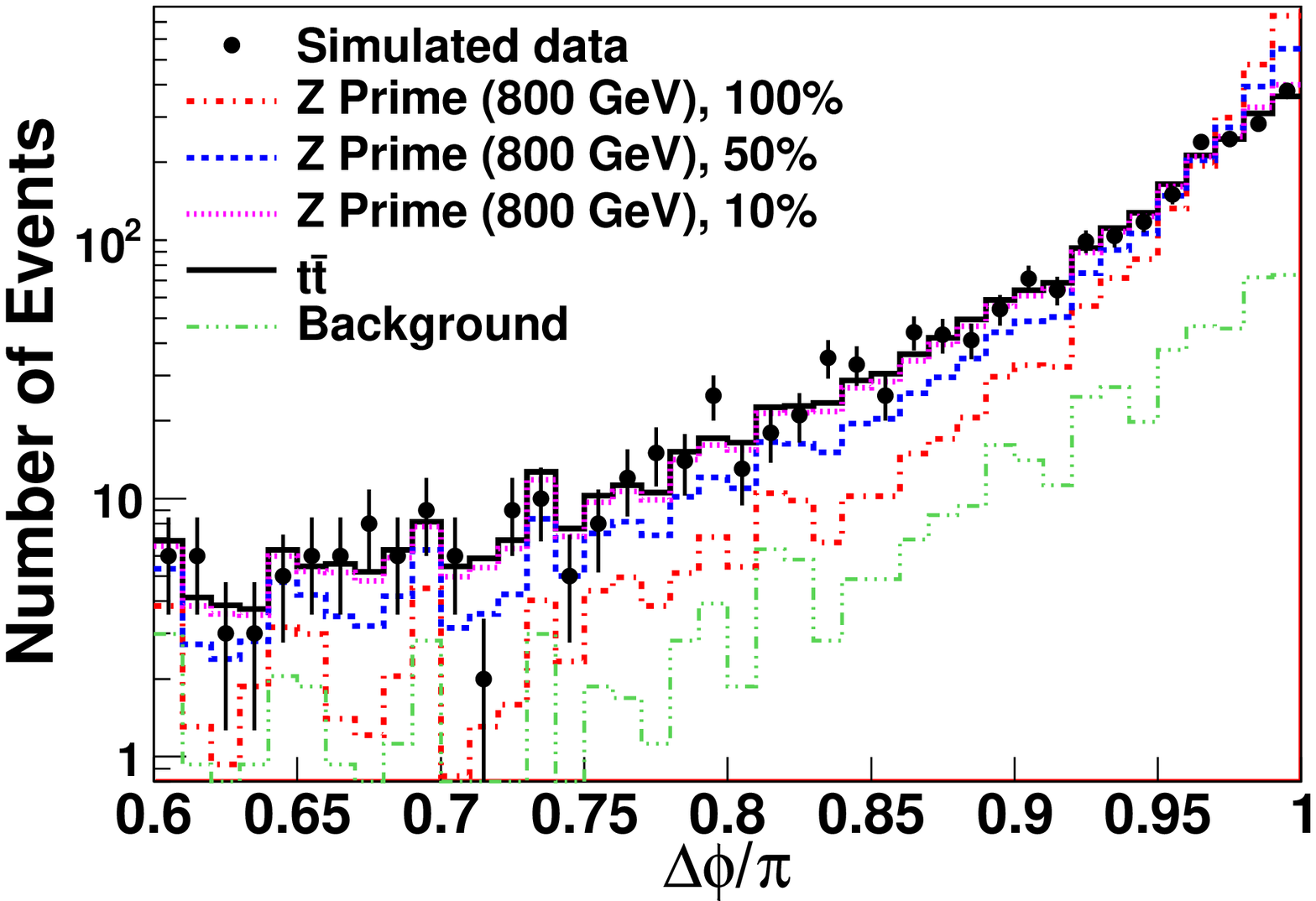} &
\includegraphics[width=0.43\textwidth]{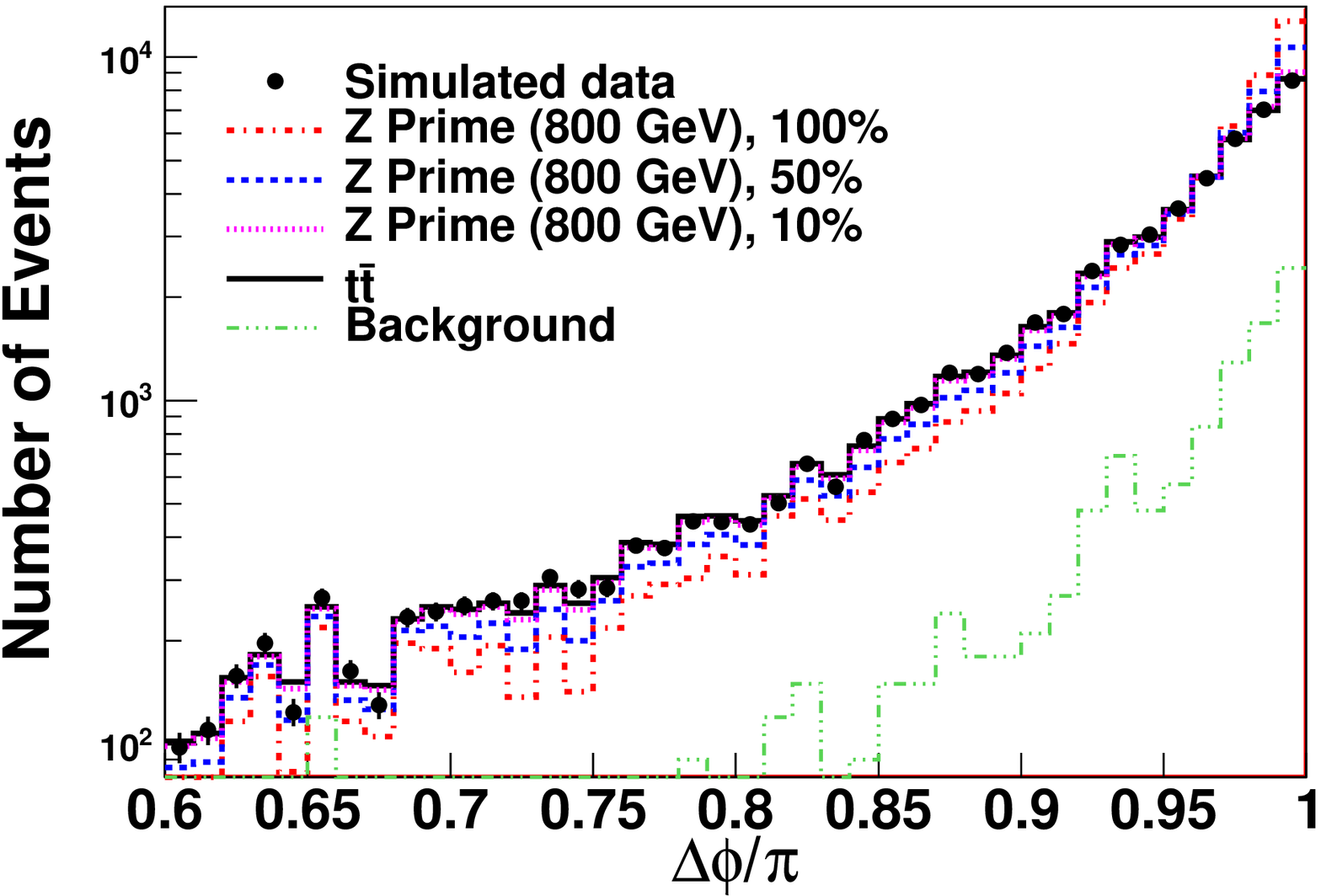} \\
(c) Tevatron (Log scale) & (d) LHC (Log scale) \\
\end{tabular}
\caption[Data fit]{(Color online) Data (simulated) compared with the 800-GeV Z$'$ model at the Tevatron and LHC with detector simulation and event reconstruction.}
\label{ref:reco1}
\end{cfigure1c}

\begin{cfigure1c}
\begin{tabular}{cc}
\includegraphics[width=0.43\textwidth]{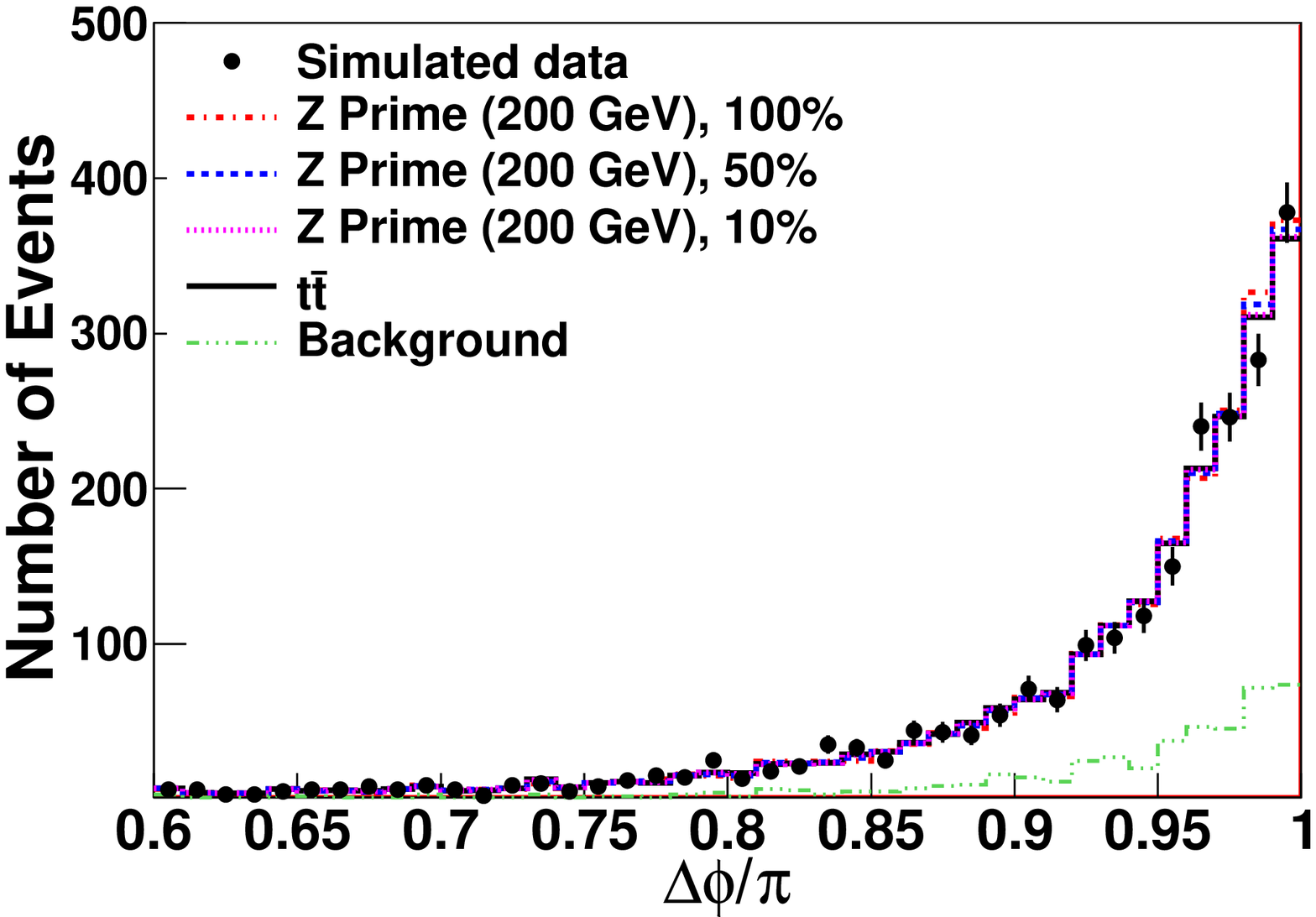} &
\includegraphics[width=0.43\textwidth]{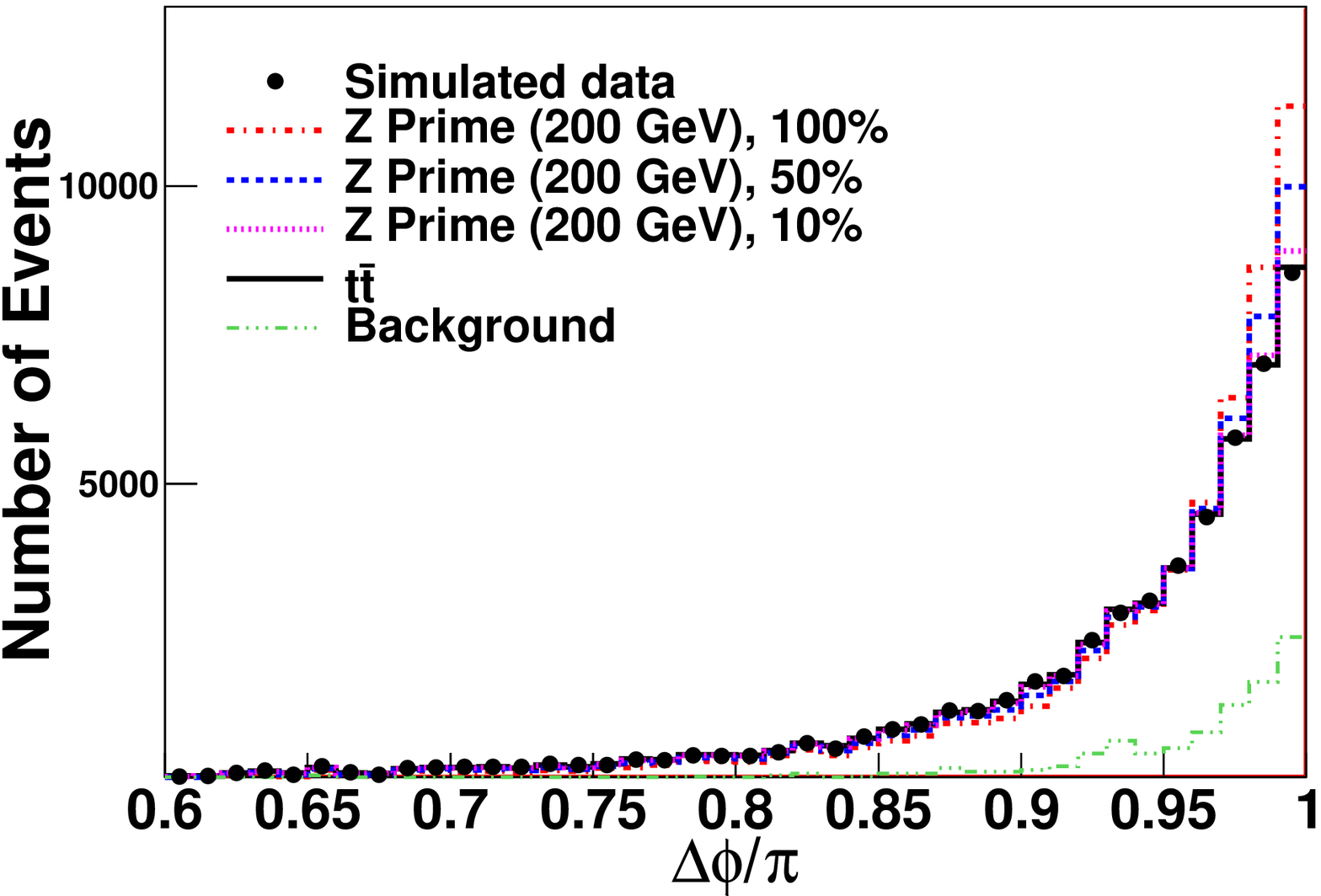} \\
(a) Tevatron  & (b) LHC  \\
\includegraphics[width=0.43\textwidth]{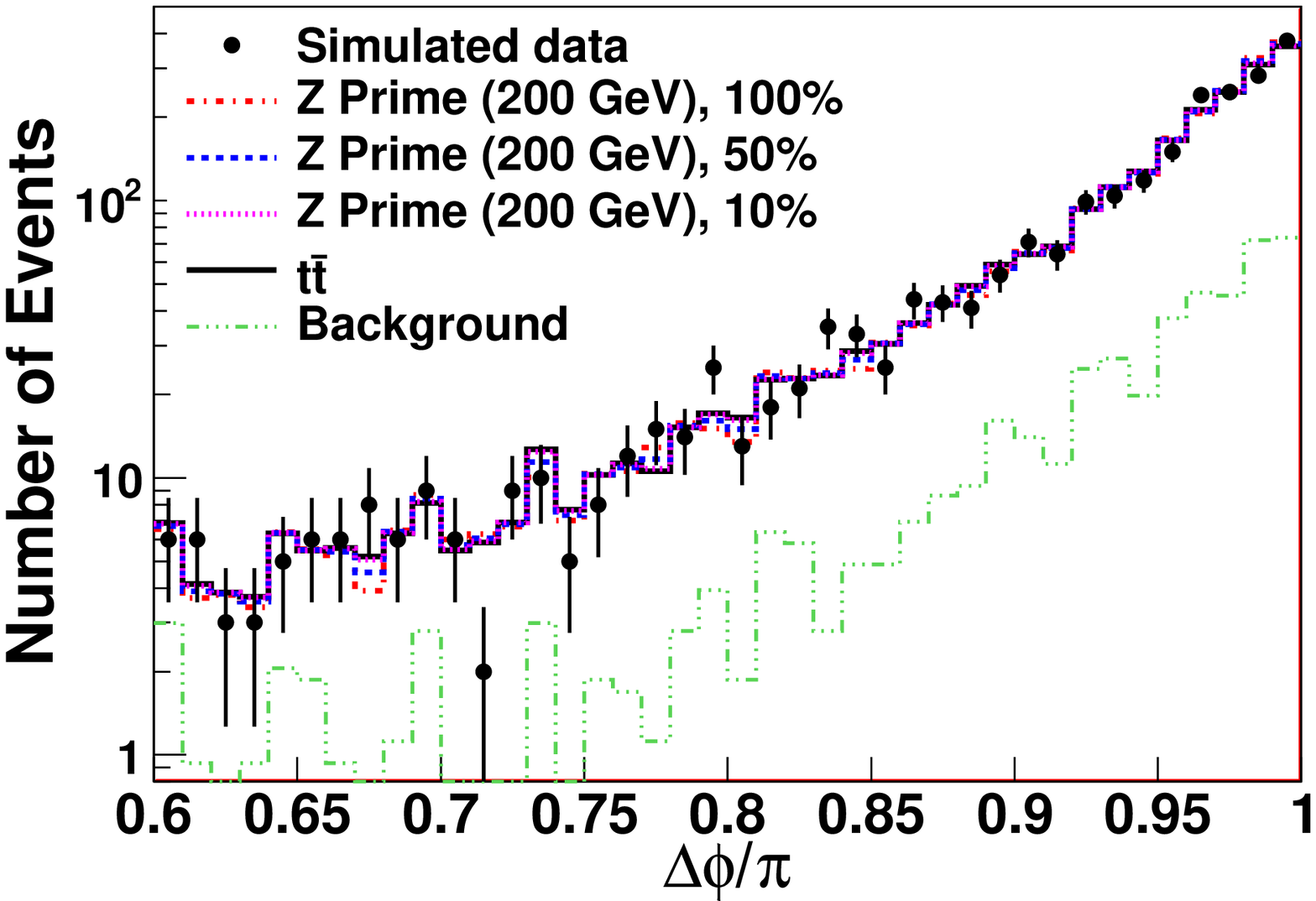} &
\includegraphics[width=0.43\textwidth]{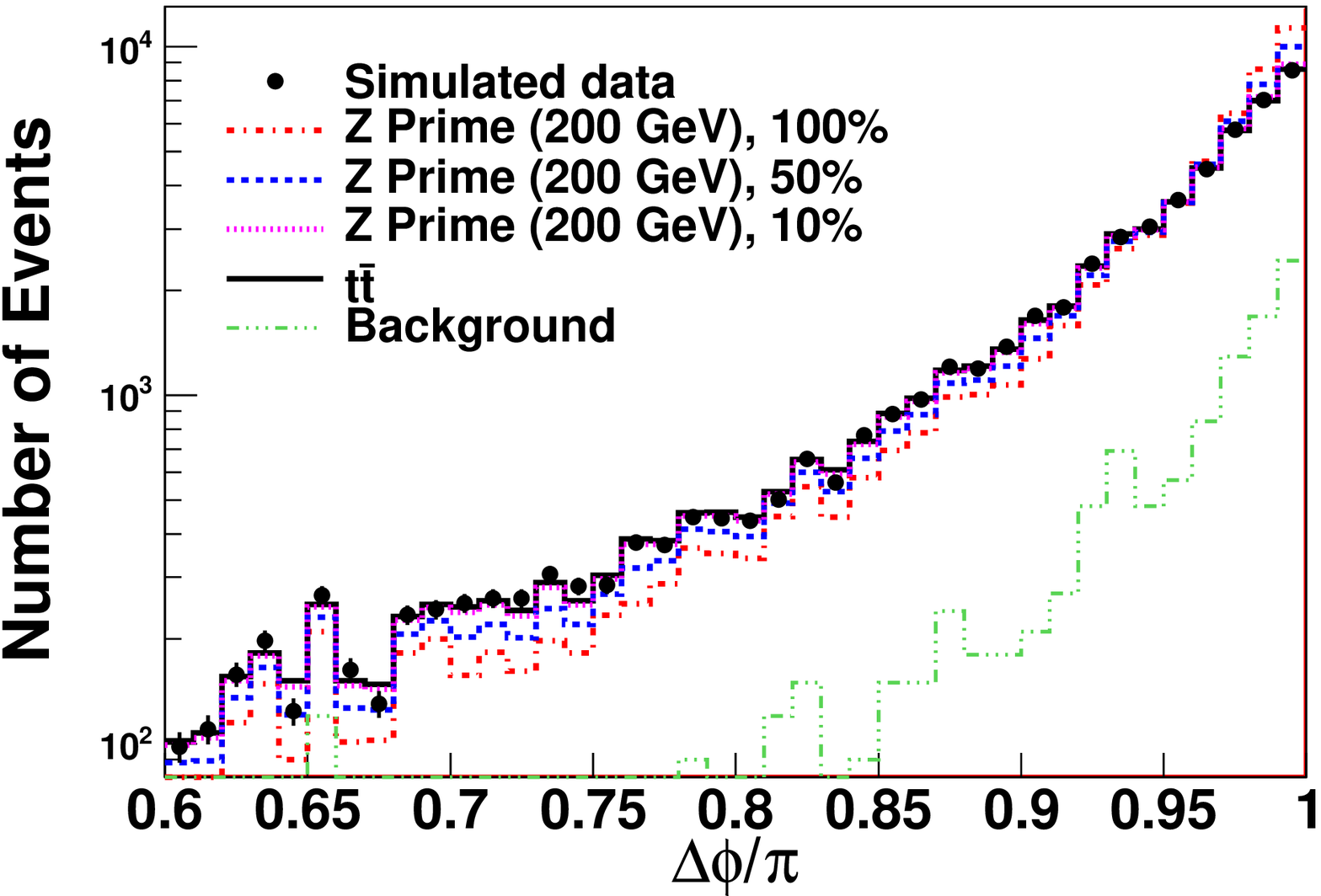} \\
(c) Tevatron (Log Scale) & (d) LHC (Log Scale) \\
\end{tabular}
\caption[Data fit]{(Color online) Data (simulated) compared with the 200-GeV Z$'$ model at the Tevatron and LHC with detector simulation and event reconstruction.}
\label{ref:reco2}
\end{cfigure1c}

\begin{cfigure1c}
\begin{tabular}{cc}
\includegraphics[width=0.43\textwidth]{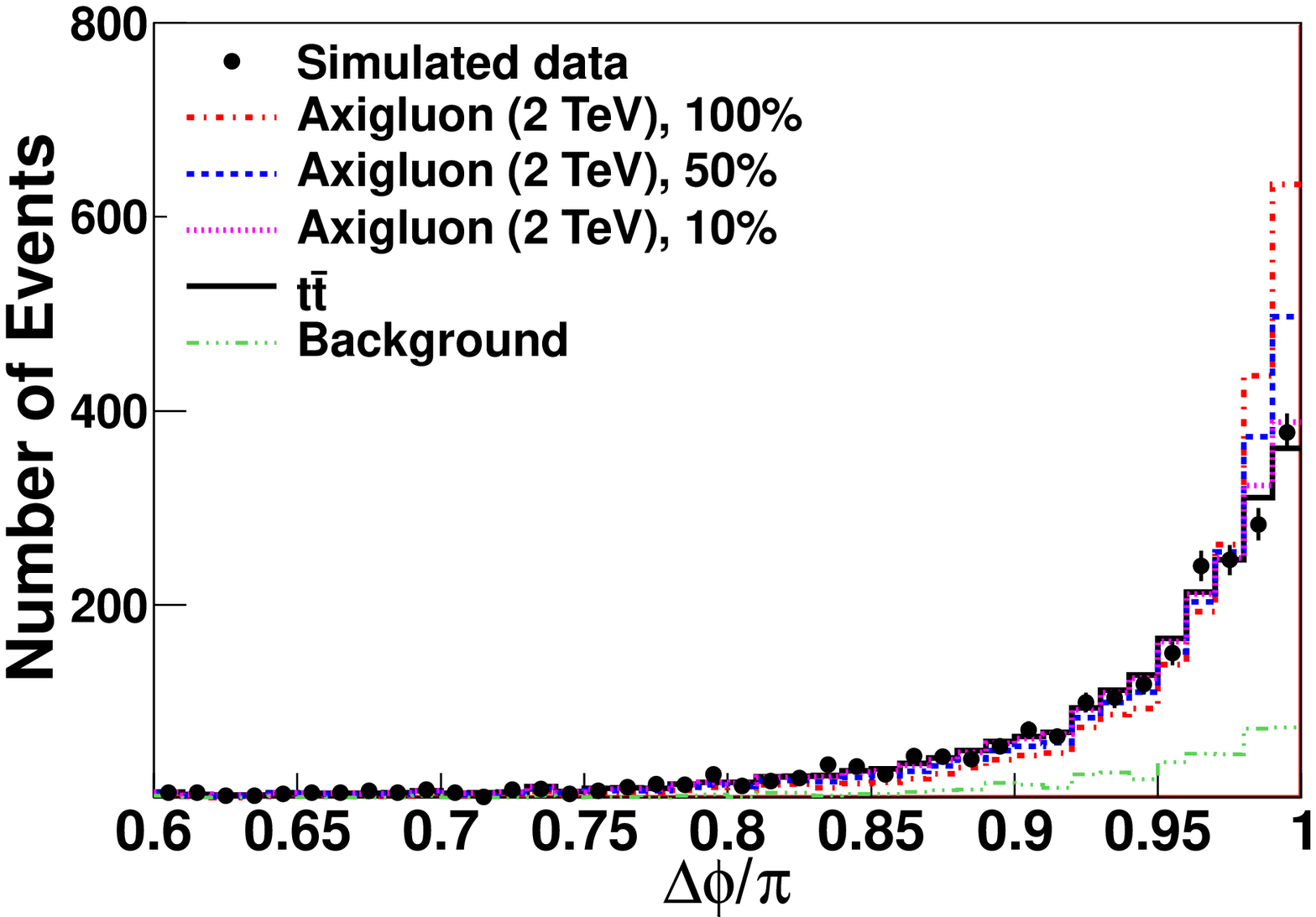} &
\includegraphics[width=0.43\textwidth]{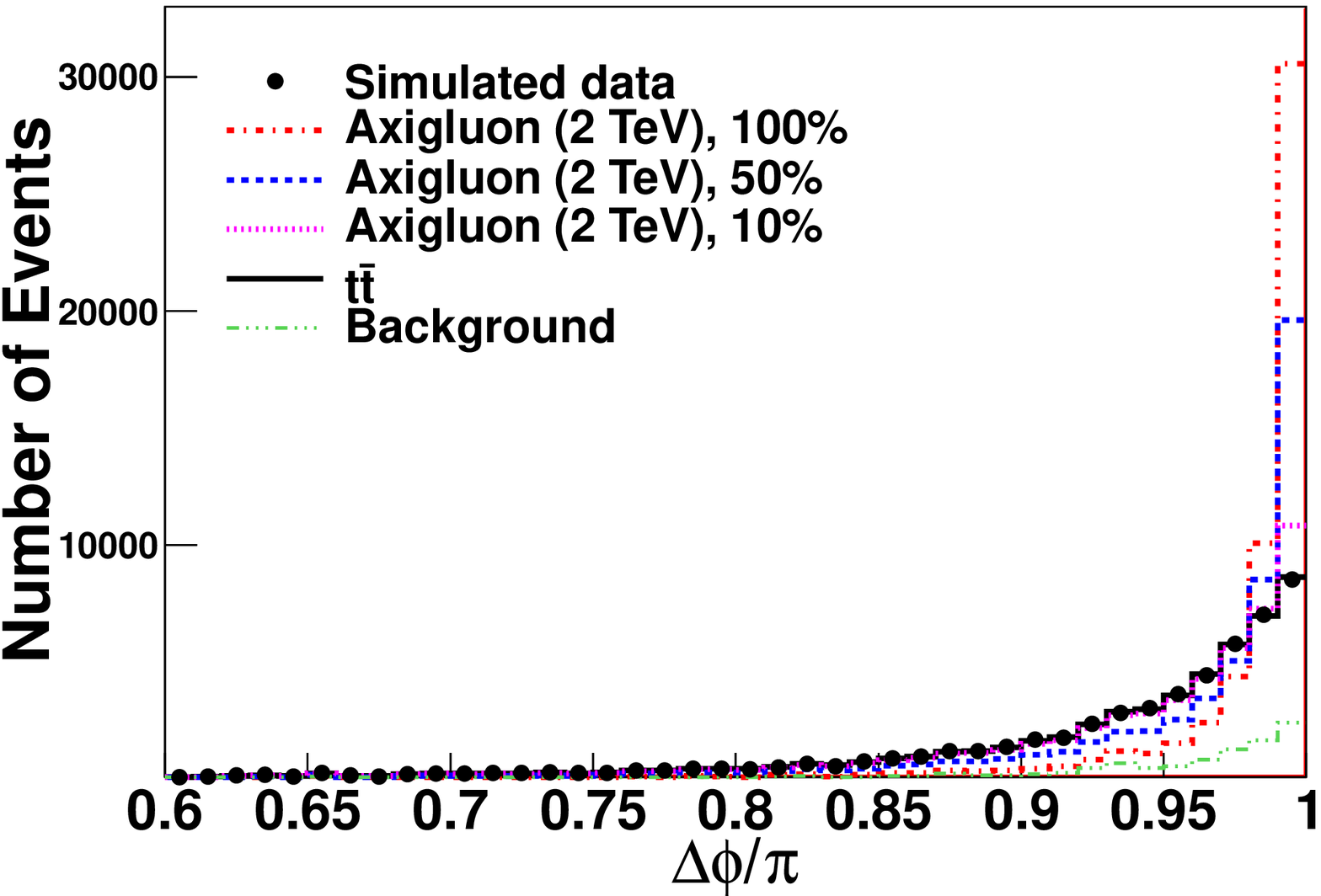} \\
(a) Tevatron & (b) LHC \\
\includegraphics[width=0.43\textwidth]{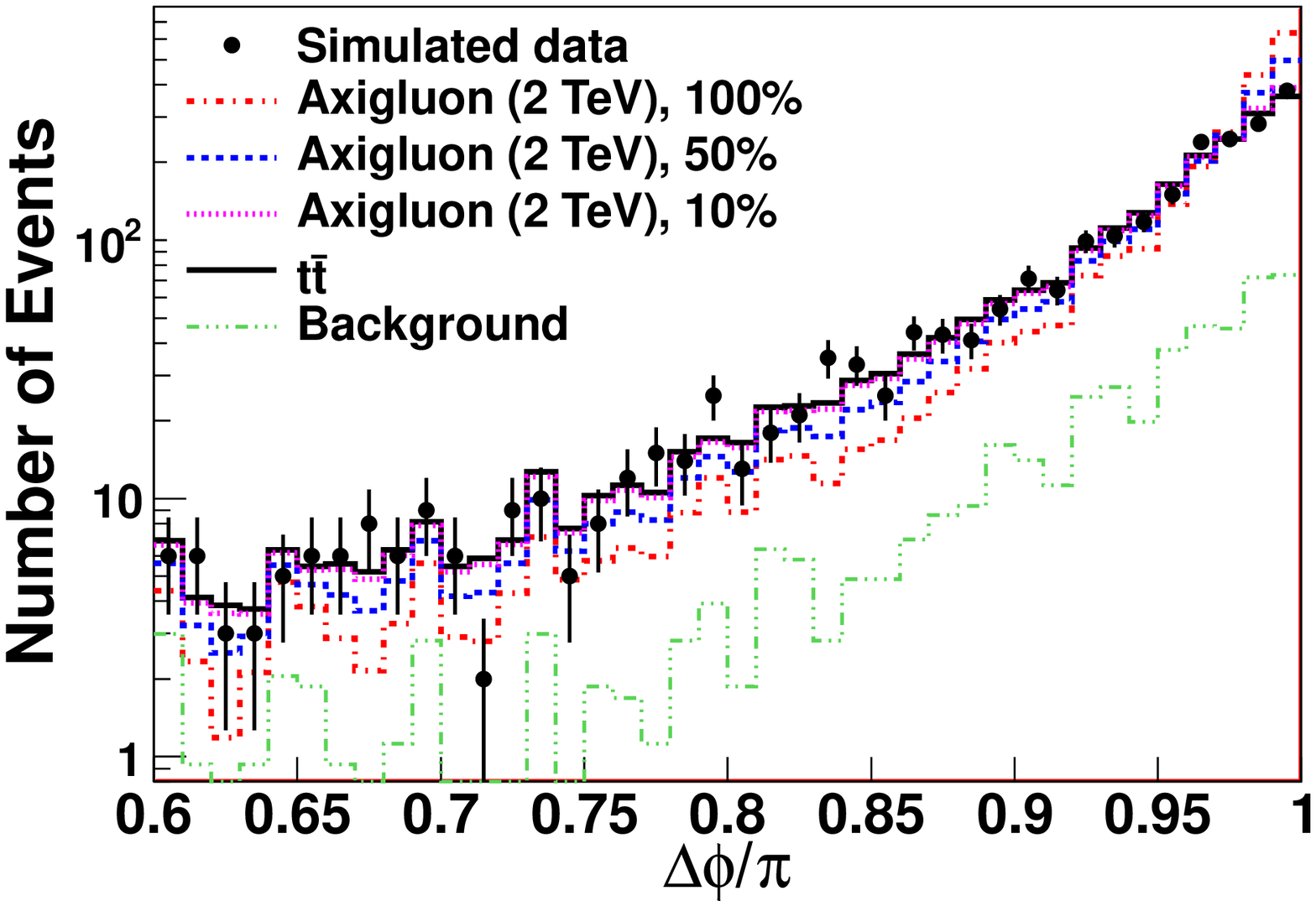} &
\includegraphics[width=0.43\textwidth]{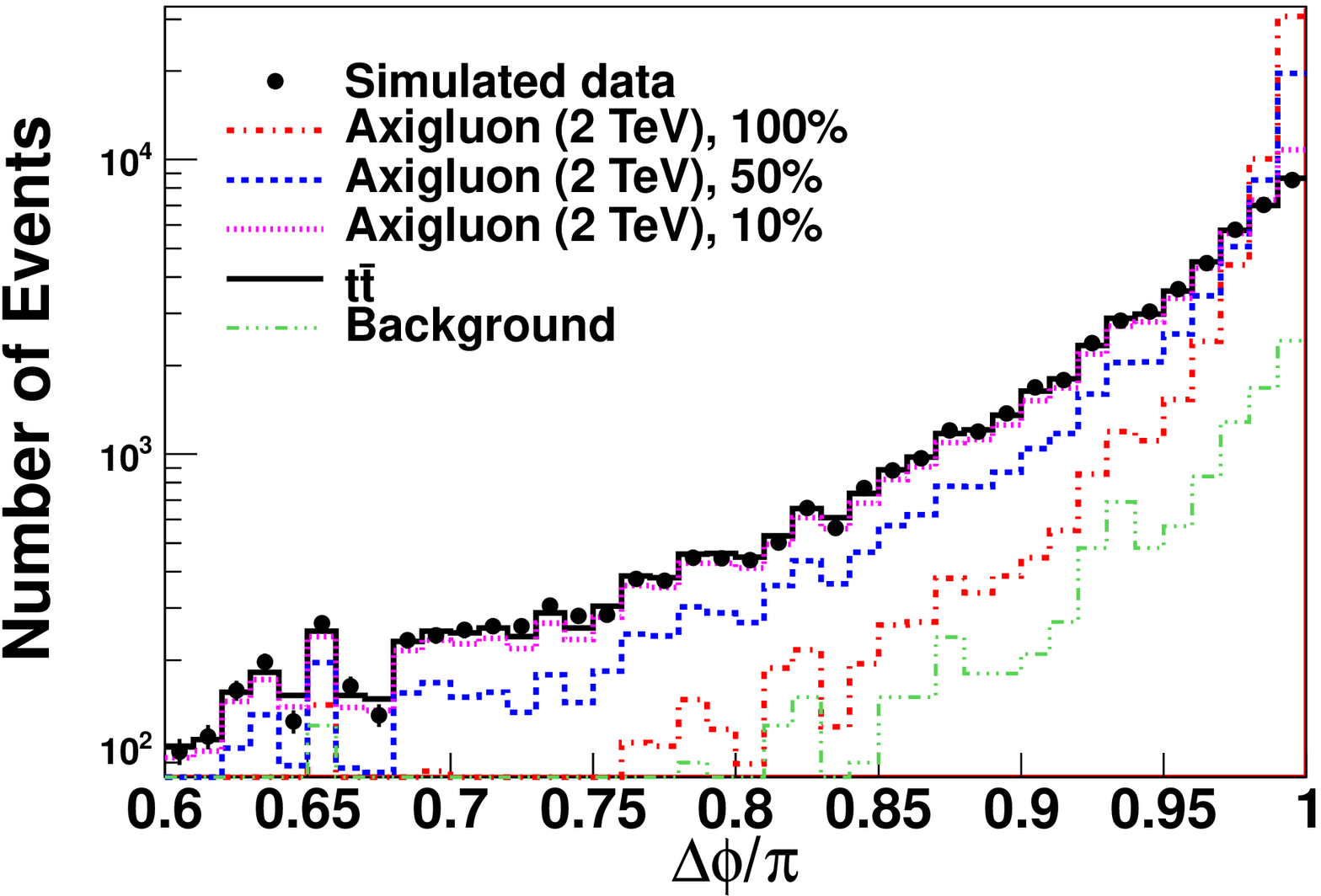} \\
(c) Tevatron (Log scale) & (d) LHC (Log scale) \\
\end{tabular}
\caption[Data fit]{(Color online) Data (simulated) compared with the 2-TeV axigluon model at the Tevatron and LHC with detector simulation and event reconstruction.}
\label{ref:reco3}
\end{cfigure1c}

To demonstrate the expected distribution of data from each collider experiment, we randomly select the signal and background events from SM \ttbar and $W$+jets background corresponding to the expected signal and background events in Table~\ref{table_background}. We call these events the ``simulated data.'' 
The $\Delta \phi$ distribution from various new physics models are compared with SM $\ttbar$ signature in Fig.s~\ref{ref:reco1},~\ref{ref:reco2}, and~\ref{ref:reco3} for the 800-GeV Z$'$, the 200-GeV Z$'$, and the 2-TeV axigluon models, respectively. Because the cross sections of new physics processes are highly depending on models, we consider three different composition of the new physics processes into the QCD process as 100~\%, 50~\%, 10~\% fractions. We consider the total events are unchanged because the expected events from the SM process are consistent with the observed events. As one can see in these figures, most of the models can be distinguished from the SM process in case of 100~\% new physics case. With \invfb{10} Tevatron data, it is clear that we can perform interesting studies for the 800-GeV Z$'$ and 2-TeV axigluon models using the $\Delta \phi$ distribution. Depending on the fraction of new physics production in the \ttbar signature, we may be able to obtain significant hints related to new physics models. However, it seems difficult to study low-mass vector boson (200-GeV Z$'$) production using $\Delta \phi$ at the Tevatron since the shape is very similar to that of the SM and the event statistics are inadequate to discriminate between them. At the LHC, there are huge differences between the shapes of the $\Delta \phi$ distributions even with event reconstruction, not only between the shapes of the new physics models and the SM but also among those of the new physics models themselves. It is possible to detect a small contribution~(less than 10\% fraction) of the new physics signature under the dominant SM \ttbar processes. Depending on the new physics cross section, the underlying physics model related to the charge forward-backward asymmetry can be studied with the $\Delta \phi$ observable.
Because $\Delta \phi$ information is not fully correlated with the charge forward-backward asymmetry, $\Delta \phi$ can give additional information about the new physics signature.

\section{Conclusion}
In conclusion, we propose an interesting new observable, $\Delta \phi$, the azimuthal decorrelation between $t$ and $\bar{t}$ quarks in  \ttbar pair production, at  hadron colliders as an interesting probe to study the SM precisely as well as to search for new physics related to the forward-backward charge asymmetry at the Tevatron. With 10 fb$^{-1}$  \ppbar collisions at the Tevatron, the possibilities for studying SM processes as well as searching for new physics signature are demonstrated. In the LHC, the large number  of \ttbar production events using \invfb{5} $pp$ collisions allows a precision study of the SM QCD processes as well as offers a huge discovery potential for new physics signatures. Together with the charge forward-backward asymmetry information, the $\Delta \phi$ observable can be used to discriminate underlying new physics theories.

\begin{acknowledgments}
SC was supported by the National Research Foundation of Korea~(NRF) Grant No. NRF-2011-0016554 and HSL was supported by NRF Grant No. NRF-2011-35B-C00007, funded by the Korean Government~(MEST).
\end{acknowledgments}

\end{document}